\documentclass[12pt]{article}
\renewcommand{\title}[1]{\begin{center}\bf\Large #1\end{center}}
\renewcommand{\author}[1]{\begin{center}\large #1\end{center}}
\usepackage{latexsym,amsfonts}
\newcommand{\rr}{\mathbb{R}}

\def\theequation{\arabic{section}.\arabic{equation}}

\begin{document}

\hspace{12.6cm}{\bf HU-EP-05/29}

\vspace{5mm}

\title{Oscillator quantization of the massive scalar
particle dynamics on AdS spacetime}

\vspace{5mm}

\author{
Harald Dorn ${}^a$
and George Jorjadze ${}^b$\\
{\small${}^a$Institut f\"ur Physik der
Humboldt-Universit\"at zu Berlin,}\\
{\small Newtonstra{\ss}e 15, D-12489 Berlin, Germany}\\
{\small${}^b$Razmadze Mathematical Institute,}\\
{\small M.Aleksidze 1, 0193, Tbilisi, Georgia}}

\vspace{9mm}

\begin{abstract}
\noindent
The set of trajectories for massive spinless particles on $AdS_{N+1}$
spacetime is described by the dynamical integrals related to the isometry
group $SO(2,N)$. The space of dynamical integrals is mapped one to one
to the phase space of the $N$-dimensional oscillator. 
Quantizing the system canonically, the classical expressions for the
symmetry
generators are deformed in a consistent way to preserve  the $so(2,N)$
commutation relations. This quantization thus yields new explicit
realizations of the spin zero positive energy UIR's of $SO(2,N)$ for generic $N$.
The representations as usual can be characterized by their minimal energy
$\alpha$ and are valid in the whole range of $\alpha$ allowed by
unitarity.  

\vspace{0.4cm}

\noindent
{\it Keywords:} AdS space; $SO(2,N)$ group; geometric and canonical
quantization;

\noindent
{\it PACS:}~~~\, 11.10. EF; 11.10. Kk; 11.10. Lm; 11.25. Hf
\end{abstract}

\baselineskip=20pt

\vspace{0.3cm}

\newpage

\section{Introduction}
$AdS$ spacetimes have an isometry group of the
same dimension as the corresponding Minkowski case. Therefore,  
quantization of particle dynamics within standard Hamiltonian reduction
techniques is possible. 
This has been done before for lower dimensional cases \cite{J,Ful,Luc}. 
Our present paper continues a systematic study of both the classical and 
quantum particle dynamics on $AdS_{N+1}$ for generic $N$ started in
\cite{DJ}. With these investigations we find new explicit realizations of 
UIR's of $SO(2,N)$. 

Another motivation for this program comes from 
possible interrelations with the corresponding quantum field theories and
string theories on 
$AdS_{N+1}$, which play a crucial role in the $AdS$/CFT correspondence, see 
e.g. \cite{ADSCFT,HF,DSS}. In this context one of the most challenging
open problems is the quantization of strings on $AdS_5\times S^5$. Certainly,
the full understanding of quantized particle dynamics on such backgrounds
could be a useful warm-up. 

There are well-known explicit expressions for field theoretical propagators
on $AdS$ spacetime. They are crucial tools for the use of the $AdS$/CFT
correspondence in application to the large $N$ and strong 't Hooft
coupling limit on  the gauge field theory side. Less is known about explicit
expressions for  field theoretical propagators on $AdS\times S$ backgrounds
\cite{DSS}.
Since they could be very helpful to understand the fate of the holographic
picture in the BMN limit \cite{bmn}, any technique for constructing propagators
should be of interest. In \cite{DJ} we have explicitely shown that at least some of the
standard $AdS$ propagators can be obtained as propagation kernels of  the
quantized particle.

Of particular interest is also the case of massless particles \cite{DJ3}, which in field
theory have been related to singleton representations of the conformal
group. For them it is natural to relate the original spacetime to the
conformal boundary of a spacetime with just one more dimension.    
 
In \cite{DJ}, quantizing the dynamics of a massive scalar particle in
$AdS_{N+1}$ space-time along the lines of geometric quantization, we have 
constructed the following representations of the isometry group generating 
$so(2,N)$ algebra  
\begin{eqnarray}
\label{repr}
&& E=\alpha +
\zeta_l\,\partial_{\zeta_l}~,~~~~~~~~~~~~~~~~~~~~~~~~~~~~~~~
J_{mn}=i(\zeta_n\,\partial_{\zeta_m}-\zeta_m\,\partial_{\zeta_n})~,~~~\\
~~~~~~~~~~~~\nonumber
&& z^*_n=2\alpha\zeta_n+(2\zeta_n\zeta_l-\zeta^2\,\delta_{nl})\,
\partial_{\zeta_l}~,~~~~~~~
 z_n=\partial_{\zeta_n}~,~~~(l,m,n=1,...,N)~.~~~
\end{eqnarray}
Here $ E$ and $ J_{mn}$ are the generators of the compact subgroup
$SO(2)\times SO(N)$, while $ z^*_n$ and $ z_n$ stand for the two kind 
of boosts of the theory.
The representations (\ref{repr}) are
characterized by the parameter $\alpha$, which corresponds to the lowest 
eigenvalue  
of the energy operator $E$. The Hilbert space is spanned by the
holomorphic functions of $N$ complex variables $\zeta_n$ $(n=1,...,N)$  
defined inside some bounded domain and the
measure for the scalar product depends on the parameter $\alpha$ and the
dimension $N$. The case $N=1$ corresponds to Bargmann's representation of
the $so(2,1)$ algebra  given on holomorphic functions inside the unit 
disk $|\zeta|<1$ (see e.g. \cite{K}), and the regularity
of the scalar product requires $\alpha>1/2$. 
For $N>1$ the integration domain is more complicated and the integration 
measure turns out to be regular for $\alpha>N-1$ only. 

Since the constructed Hilbert space seems to have some similarities to the 
Fock space of a $N$-dimensional oscillator, it is natural to ask for a
corresponding explicit construction in terms of oscillator variables
and its canonical quantization. From  \cite{DJ} there remained also 
another puzzling question. The allowed values of $\alpha$ do not exhaust
the well-known unitarity bound, see e.g. \cite{Min} and refs. therein,
\begin{equation}
\label{UB}
\alpha\geq (N-2)/2~~~\mbox{for}~ N\geq 2~~~\mbox{ and}~~~
\alpha \geq 0~~~\mbox{for}~ N=1~.
\end{equation}
In addition, within the representations 
(\ref{repr}) one cannot reach values for $\alpha$, corresponding 
to the Casimir number $C=(1-N^2)/4$ related to a conformal
invariant setting in $AdS_{N+1}$.

The formulation in terms of oscillator variables for the simplest case $N=1$
can be obtained by the Holstein-Primakoff method \cite{GP} in the form 
\begin{equation}
\label{a,a^*}
 z=\sqrt{H+2\alpha}~\, a~, ~~~~~ z^*=
 a^*\sqrt{H+2\alpha}~,~~~~~~~~~  E= H+\alpha~,
\end{equation}
where $H$ is the normal ordered oscillator Hamiltonian $H =a^* a$ 
and the square root of the positive operator
$H+2\alpha$ is defined in standard manner via spectral representation. 
The representation (\ref{a,a^*}) is unitary and irreducible for
$\alpha>0$. 
Thus, the canonical scheme covers all lowest weight UIR's of the $so(2,1)$ 
algebra. The aim of the present paper
is to generalize this result to arbitrary $N$.

\setcounter{equation}{0}
\section{Classical theory}

The $N+1$ dimensional $AdS$ space can be represented as the universal
covering of the hyperboloid
of radius $R$
\begin{equation}\label{hyp}
 X_0^2+X_{0'}^2-\sum_{n=1}^NX_n^2=R^2
\end{equation}
embedded in the $N+2$ dimensional flat space $\rr_N^2$
with coordinates $X_A,$ $A=(0,0',1,...,N)$
and the metric tensor $\,G_{AB}=\mbox{diag}(+,+,-,...,-)$.
The induced metric tensor on the hyperboloid
has Lorentzian signature and the polar 
angle $\theta$ in the $(X_0, X_{0'})$ plane can be taken as the time
coordinate:
$X_{0}=r\cos\theta$, 
$X_{0'}=r\sin\theta~$ ($r^2=X_nX_n+R^2$).

The dynamics of a massive particle moving on
the hyperboloid we describe by the action 
\begin{equation}\label{S}
S=-\int d\tau \left[m\sqrt{\dot X^A\dot X_A}+
{\frac{\mu}{2}}(X^AX_A-R^2)\right]~,
\end{equation}
where $m>0$ is the particle mass,  $\mu$ is a Lagrange multiplier and 
$\tau$ is an evolution parameter. 
The time direction we fix in a $SO(2,N)$-invariant way
by requiring 
$\dot\theta>0$, which is equivalent to ${X_0\dot X_{0'}-X_{0'}\dot X_0}>0.$
The space-time isometry group $SO(2,N)$ 
provides the Noether  conserved quantities
\begin{equation}\label{M_AB}
J_{AB} =P_A\,X_B -P_B\,X_A~,
\end{equation}
where $P_A$ are the canonical momenta. 
Since $\theta$ is the time
coordinate, $J_{00'}$ is associated with the particle energy. We denote it by 
$E$ and due to the choice of the time direction it is positive
\begin{equation}\label{E}
E=P_0X_{0'}-P_{0'}X_0= \frac{m}{\sqrt{\dot X^A\dot X_A}}
(X_0\dot X_{0'}-X_{0'}\dot X_0)>0~.
\end{equation}
The conservation of $J_{AB}$ allows to represent the set of
trajectories geometrically without solving the dynamical
equations. From Eq. (\ref{M_AB}) we find N equations as identities in
the  variables $(P,X)$
\begin{equation}\label{traj}
E\,X_n=J_{0n}\,X_{0'}-J_{0'n}\,X_0~,~~~~~~~~(n=1,...,N)~.
\end{equation}
Since $E$, $J_{0'n}$, $J_{0'n}$ are constants, Eq. (\ref{traj})
defines a 2-dimensional plane in the embedding space $\rr_N^2$ and
the intersection of this plane with the hyperbola (\ref{hyp}) is a
particle trajectory. 

The action (\ref{S}) is invariant under the reparametrizations
$\tau\rightarrow f(\tau)$ with the corresponding transformations
of the Lagrange multiplier ($\mu \rightarrow \mu\,/ f'$). 
The gauge symmetry, as usual, leads to the
dynamical constraints. Applying Dirac's procedure, we
find three constraints
\begin{equation}\label{Phi=0}
X^AX_A-R^2=0~,~~~~~~~~P_AP^A-m^2=0~,~~~~~~~~~ P_A\,X^A=0~,
\end{equation}
which fix the quadratic Casimir number of the symmetry group
\begin{equation}\label{C}
C=\frac{1}{2}\, J_{AB}J^{AB}=m^2R^2~.
\end{equation}
For further calculations it is convenient to introduce complex 
valued dynamical integrals
\begin{equation}\label{z_n}
z_n=J_{0'n}-iJ_{0n}~,~~~~~~~~~~z_n^*=J_{0'n}+iJ_{0n}~,
\end{equation}
and the  scalar variables $\lambda^2=z^*_nz_n,$
$\rho^4={z^*}^2\,z^2$ (with $z^2=z_nz_n$). Then (\ref{C}) becomes
\begin{equation}\label{C1}
E^2+J^2=\lambda^2+\alpha^2~,
\end{equation}
where
\begin{equation}\label{alpha,M}
J^2={\frac{1}{2}\,
J_{mn}J_{mn}}~~~~~~~\mbox{and}~~~~~~~\alpha=mR~.
\end{equation}
A set of other quadratic relations follows from (\ref{M_AB}) as
identities in the variables $(P, X)$
\begin{equation}\label{MM=MM}
J_{AB}\,J_{A'B'}= J_{AA'}\,J_{BB'}-J_{AB'}\,J_{BA'}~.
\end{equation}
These equations are nontrivial in terms of the dynamical
integrals, if all indices $A$, $B$, $A'$, $B'$ are different.
Taking $A=0$, $B=0'$, $A'=m$ and $B'=n$ $(m\neq n)$ we obtain
\begin{equation}\label{EM=zz}
2iE\,J_{mn}=z_m^*z_n-z_n^*z_m~,
\end{equation}
and its square yields  $4E^2J^2=\lambda^4-\rho^4$. Then, together 
with (\ref{C1}) 
we conclude that $E^2$ and $J^2$ are roots of the
quadratic equation $\,4x^2-4(\lambda^2+\alpha^2)x+\lambda^4-\rho^4=0$
and find 
\begin{eqnarray}\label{E=}
E^2=\frac{1}{2}\,\left(\lambda^2+\alpha^2+
\sqrt{\alpha^4+2\alpha^2\lambda^2+\rho^4}\,\right)~.
\end{eqnarray}
We neglect the second root as unphysical, since it does not provide positivity
of the energy (see \cite{DJ} for more details). 
According to (\ref{E=}) $\alpha$ is the lowest value of energy. 
Eqs. (\ref{E=}) and (\ref{EM=zz}) define $E$ and $J_{mn}$ as functions
of ($z_n, z_n^*$) and, therefore, ($z_n, z_n^*$) are global coordinates on the 
space of dynamical integrals, which is the physical phase space 
$\Gamma_{ph}$ of the system.

Based on the canonical brackets between the variables $(P,X)$ 
the $so(2,N)$ Poisson bracket algebra of the symmetry generators can be
written as
\begin{eqnarray}\label{PB_z}
&&\{z_m^*,\,z_n\}=2J_{mn}-2i\delta_{mn}\,E~,~~~~~
\{z_m,\,z_n\}=0=\{z_m^*,\,z_n^*\}~,\\\label{PBE,J}
&&\{J_{lm},\,z_n\}=z_l\,\delta_{mn}-z_m\,\delta_{ln}~,
~~~~~~\{E,\,z_n\}=-iz_n~,~~~~~ \{E,\,J_{mn}\}=0~,
\end{eqnarray}
and $J_{mn}$'s form the $so(N)$ algebra.
These relations are preserved after reduction to $\Gamma_{ph}$, 
since the constraints (\ref{Phi=0}) are $SO(2,N)$ scalars. 
But (\ref{PB_z})-(\ref{PBE,J})
 become essentially non-linear in terms of the
independent variables $z_n$, $z_n^*$ and their quantum realization  
is a nontrivial issue. 

Our aim is to find a canonical parametrization of $\Gamma_{ph}$ and then
to quantize the system canonically.
For this purpose we introduce a set of $2N$ variables
($a_n$, $a_n^*;\, n=1,...,N$) with the canonical Poisson brackets 
\begin{equation}\label{C-PB}
\{a_n,a_m\}=0=\{a_n^*,a_m^*\}~,~~~~~~~~~~~~
\{a_n,a_m^*\}=i\delta_{mn}~,
\end{equation}
and assume that the generators of the compact
transformations are given by
\begin{eqnarray}\label{E, J_mn} 
E=H+\alpha~,~~~~~~~~J_{mn}=i(a_n^*a_m-a_m^*a_n)~,
\end{eqnarray}
where $H=a_n^*a_n$ is the oscillator Hamiltonian. Note that Eqs. 
(\ref{C-PB})-(\ref{E, J_mn})
immediately provide the $so(2)\times so(N)$ part of $so(2,N)$ algebra
and the $so(N)$ scalar (\ref{alpha,M}) 
constructed from $J_{mn}$ becomes $J^2=H^2-{a^*}^2a^2$, 
with ${a}^2=a_na_n$ and ${a^*}^2=a^*_na^*_n$.

We look now for a parametrization of the  generators $z_n$ and $z_n^*$ in
the form
\begin{equation}\label{z_n=XY?}
z_n=X(H,J^2)\,a_n+Y(H,J^2)\,a^2\,a^*_n~,~~~~~~~~~z^*_n=X(H,J^2)\,a^*_n
+ Y(H,J^2)\,{a^*}^2\,a_n~,
\end{equation}
where  $X$ and $Y$ are real functions of the scalar variables ($H,\,J^2)$.
This ansatz guarantees the correct commutation relations between compact and
non-compact generators. 
The Casimir condition (\ref{C1})
and the quadratic relation (\ref{EM=zz}) yield two equations for
the functions $X$, $Y$
\begin{eqnarray}\label{X,Y=?}
&&HX^2+(H^2-J^2)(2XY +HY^2)=H^2+J^2+2\alpha H~,\\ \nonumber
&&X^2-(H^2-J^2)Y^2=2(H+\alpha)~.
\end{eqnarray}
After elimination of $X$, (\ref{X,Y=?}) reduces to a quadratic
equation for $Y^2$. Choosing the root which is nonsingular for
$J^2=0$ and putting $Y=-\sqrt{Y^2}$, we obtain
\begin{eqnarray}\label{Y=}
X=-\frac{1}{2Y}\,[1+2H\,Y^2]~,~~~~~
Y=-\left(2H+4\alpha+2\sqrt{(H+2\alpha)^2
-J^2}\right)^{-\frac{1}{2}}~.
\end{eqnarray}
Note that the choice $Y=\sqrt{Y^2}$ leads to a canonically
equivalent answer, reproduced from (\ref{z_n=XY?}) by the inversion
$a_n\mapsto -a_n$.

Due to (\ref{z_n=XY?}) and (\ref{Y=}) the $SO(N)$ scalars $z^2$ and $a^2$
are related by 
\begin{equation}\label{z^2=a^2}
z^2=F\,{a}^2~,
\end{equation}
where $F$ is the following function of real scalar variables
\begin{equation}\label{F}
F=\sqrt{(H+2\alpha)^2-J^2}=\sqrt{(E+\alpha)^2-J^2}~.
\end{equation}
This function also played an important role in the scheme of geometric
quantization \cite{DJ}. Note that the
parametrization of $\Gamma_{ph}$ given by (\ref{z_n=XY?}) and (\ref{Y=})
can be written as
\begin{eqnarray}\label{z_n=}
&&z_n= \frac{1}{\sqrt{2(H+2\alpha+F)}}\left((2H+2\alpha+F)\,a_n-
{a}^2\,a^*_n\right)~.
\end{eqnarray}

The relation (\ref{z^2=a^2}) helps to invert (\ref{z_n=XY?})  and we find
\begin{equation}\label{a=z}
a_n=\frac{1}{2E}\left(X\,z_n-\frac{Y}{F}\,{z}^2\,z_n^*\right)~,
~~~~~~~~~~~~a_n^*=
\frac{1}{2E}\left(X\,z_n^*-\frac{Y}{F}\,{z^*}^2\,z_n\right)~,
\end{equation}
where  now $X$, $Y$ and $F$ are treated as functions of $(E,~J^2)$ and
they are obtained  from (\ref{Y=}) replacing
$H$ by $E-\alpha$.

The direct calculation shows that the canonical brackets
(\ref{C-PB}) lead to the $so(2,N)$ algebra
(\ref{PB_z})-(\ref{PBE,J}).
To our knowledge, the constructed canonical parametrization 
and its quantum realization for generic
$N$ was not discussed in the literature before.

\setcounter{equation}{0}

\section{Quantum theory}

We quantize the $AdS$ particle dynamics using the
canonical coordinates constructed in the previous section. 
We will use for quantum operators the
same letters as for the corresponding
classical variables. Canonical
quantization assumes a realization of the canonical
commutation relations
\begin{equation}\label{CCR}
 [ a_n, a_m]=0=[ a_n^*, a_m^*]~,~~~~~~~~~~~
 [ a_n, a_m^*]=\delta_{mn}~,
\end{equation}
and a construction of the symmetry generators  $E$, $J_{mn}$, $z_n$, $z_n^*$
on the basis of the parametrization  (\ref{E, J_mn}), 
(\ref{z_n=XY?}) and (\ref{Y=}). Due to the non-linearity of the
parametrization we face with the problem of ordering ambiguity.

There is not such a problem for the generators of $SO(N)$
rotations
\begin{equation}\label{M_mn^}
J_{mn}=i(a^*_n a_m- a^*_m a_n)~,
\end{equation}
and the operator $J^2$ in terms of the creation-annihilation
operators becomes
\begin{equation}\label{M^2^}
{J}^2=\frac{1}{2} J_{mn} J_{mn}={H}^2+(N-2)H- {a}^*\,^2{a}^2~,
\end{equation}
where $H=a^*_n a_n$ is the normal ordered oscillator Hamiltonian,
$a^2= a_n a_n$ and ${ a}^*\,^2= a^*_n a^*_n$.

Classically we had for the quadratic Casimir $C=m^2R^2=\alpha ^2$.
In the quantum case we continue to denote the lowest energy value
by $\alpha $, i.e.
\begin{equation}\label{E^}
E= H+\alpha~,
\end{equation}
but we have to expect a renormalization of the relation of $\alpha $ to the
quadratic Casimir, which defines the squared mass of the quantum particle.
Since we are interested in representations which are unitary equivalent
to those constructed by geometric quantization we take
the corresponding renormalized relation from \cite{DJ}
\begin{equation}
\label{C^}
C= E^2+\frac{1}{2}\,J_{mn}J_{mn}-\frac{1}{2}
\,(z^*_n z_n + z_n z^*_n)= \alpha(\alpha-N)~.
\end{equation}
Note in addition, that the quadratic relation (\ref{EM=zz}) due to
(\ref{repr}) is deformed into
\begin{eqnarray}
\label{EM^}
z^*_m z_n-z^*_n z_m=2i(E-1)J_{mn}~,~~~~~ \mbox{or}~~~~~~
z_n z^*_m-z_mz^*_n=2i(E+1)J_{mn}~.
\end{eqnarray}
The ordering problem is more complicated for $z_n$ and $ z^*_n$.
We look for them in the form
\begin{equation}\label{z_n^}
 z_n=X_N(H,J^2)\,a_n\, +Y_N(H, J^2)\,a^2\,a^*_n~,
 ~~~~~~~~~z^*_n=
a^*_n\,X_N(H, J^2)+ a_n\,{a^*}\,^2\,Y_N(H, J^2)~,
\end{equation}
where  $X_N$ and $Y_N$ are real unknown functions of the commuting
operators $H$ and $J^2$, similarly to (\ref{z_n=XY?}). 
Due to the ordering ambiguity, the functions
$X_N$, $Y_N$ are quantum mechanical deformations
of $X$, $Y$ and the index $N$ indicates that the deformations can
depend on the dimension $N$.
To fix these functions we use the
quantum versions of the Casimir condition (\ref{C^}) and the quadratic relation
(\ref{EM^}), which can be written as
\begin{equation}\label{z_nz_n*}
z_n\,
z^*_n=H^2+J^2+(2\alpha+N)H+2N\alpha~~~~~\mbox{and}~~~~z_n\,z^*_m
-z_m\,z^*_n  =2(H+\alpha+1)(a^*_m a_n-a^*_n a_m)~,
\end{equation}
respectively. Taking into account 
the commutation relations between the scalar
operators
\begin{equation}\label{[a^2,a*^2]}
[ H,  a^2]=-2 a^2,~~~~~~~~[ H,{ a^*}\,^2]= 2{ a^*}\,^2~,~~~~~~
[a^2,{ a^*}\,^2]=4H+2N~,
\end{equation}
from (\ref{z_nz_n*}) we find two equations for $X_N,~Y_N$
\begin{eqnarray}\label{X,Y=?^}
&&(H+N)X_N^2+\left(H^2-J^2+(N+2)H+2N\right)\left(2X_NY_N
+(H+2)Y_N^2\right)=~~~~~~~~~~~~~~\\
\nonumber
&&~~~~~~~~~~~~~~~~~~~~~~~~~~~~~~~~~~~~~~~
~~~~~~~~~~~~~~~~~H^2+J^2+(2\alpha +N)H+2\alpha N~,\\
\nonumber
&&X_N^2-\left(H^2-J^2+(N+2)H+2N\right)Y_N^2=2(H+\alpha+1)~.
\end{eqnarray}
These equations contain only commuting operators and they can be solved as
algebraic equations like (\ref{X,Y=?}) in the classical case.
After elimination of $X_N$, (\ref{X,Y=?^}) reduces to a
quadratic equation for $Y_N^2$. Neglecting the root, which
corresponds to the singular solution in the classical limit, and
choosing $\,Y_N=-\sqrt{Y_N^2}\,$, we obtain  (see (\ref {Y=}))
\begin{eqnarray}\label{Y_N=}
&&X_N=-\frac{1}{2Y_N}\,[1+(2H+N+2)\,Y_N^2]~,\\
\nonumber &&Y_N=-\left(2H+4\alpha-N+2+2\sqrt{(H+2\alpha)^2-
(N-2)(H+2\alpha)-J^2}\right)^{-\frac{1}{2}}~.
\end{eqnarray}
These operator expressions are naturally defined on the
eigenstates of $H$ and $J^2$ as multiplication operators. 
The choice $\,Y_N=\sqrt{Y_N^2}\,$ leads to an
unitary equivalent answer, since the change of sign of $Y_N$
corresponds to the inversion $a_n\rightarrow -a_n$.

Eqs. (\ref{z_n^}) and (\ref{Y_N=}) provide a deformed version of 
(\ref{z_n=})
\begin{eqnarray}\label{z_n=1}
&&
z_n=\frac{1}{\sqrt{2(H_N+2\alpha_N+F_N)}}\left((2H_N+2\alpha_N+F_N)\,a_n-
{a}^2\,a^*_n\right)~,~~~~\mbox{with}\\
\label{X,Y=}
 &&F_N=\sqrt{(H_N+2\alpha_N)^2-J_N^2}~,~~~~~~~~~~~~~~~~~~~
\alpha_N=\alpha-\frac{N}{2}~,
\\
\nonumber &&H_N=H+\frac{N+2}{2}~,~~~~~~~~~~~~~~~~~~~
J_N^2=J^2+\left(\frac{N-2}{2}\right)^2~.
\end{eqnarray}
The notation in the above formulas has been chosen in a form to make
both the structural similarities as well as the modifications 
relative to their classical counterparts
(\ref{F})-(\ref{z_n=}) manifest. 
One can check that for $N=1$ Eq. (\ref{z_n=1}) 
reproduce the operator $z$ in (\ref{a,a^*}).

The ansatz (\ref{z_n^}) obviously satisfies the commutation
relations of $\,z_n\,$ with $\,E\,$ and $\, J_{mn}\,$ for
arbitrary $X_N$ and $Y_N$
\begin{equation}\label{so}
[E, z_n]=-z_n~,~~~~~[ J_{lm},
z_n]=i(\delta_{ln}\,z_m-\delta_{mn}\,z_l)~.
\end{equation}
To calculate the commutation relations between the operators
(\ref{z_n^}) we use the exchange relations of the
creation-annihilation operators with the scalar operators. These
relations are derived in the Appendix and by
(\ref{fa_n})-(\ref{tildeY}) we find an alternative form of
(\ref{z_n^})
\begin{equation}\label{z_n=2}
z_n= a_n\,\tilde X_N + a^*_n\,a^2\,\tilde Y_N~, ~~~~z^*_n=\tilde
X_N\,a^*_n+ \tilde Y_N\,{a^*}\,^2\,a_n~,~~~~\mbox{where}
\end{equation}
\begin{eqnarray}\label{X',Y'}
&&\tilde X_N= -\frac{1}{2\tilde Y_N}\left(1+(2H+N-2)\,\tilde
Y_N^2\right)~,\\ \nonumber &&\tilde Y_N=-\left(2H+4\alpha-N-2+2
\sqrt{(H+2\alpha)^2-(N+2)(H+2\alpha)+2N-J^2}\right)^{-\frac{1}{2}}~.
\end{eqnarray}
Note that the functions (\ref{Y_N=}) and (\ref{X',Y'}) are related
by
\begin{eqnarray}\label{X,Y=1}
\tilde X_N(H,J^2)= X_N(H-2,J^2)~,~~~~~~~~~\tilde Y_N(H,J^2)=
Y_N(H-2,J^2)~.
\end{eqnarray}
The operators (\ref{z_n=2})-(\ref{X',Y'}) satisfy the equations
\begin{equation}\label{z_nz_n*1}
z^*_n\,z_n=H^2+J^2+(2\alpha-N)H~~~~~\mbox{and}~~~~z^*_m z_n-z^*_n
z_m =2(H+\alpha-1)(a^*_m a_n-a^*_n a_m)~,
\end{equation}
which are equivalent to (\ref{C^}) and (\ref{EM^}), respectively. 

It remains to 
check the commutation relations
\begin{equation}\label{so1}
[z_m, z_n]=0~,~~~~~[ z_m^*, z_n]=2(a_n^*a_m-a_m^*a_n)
-2\delta_{mn}\,(H+\alpha)~.
\end{equation}
Using the two representations of $z_n$ operators (\ref{z_n^}) and
(\ref{z_n=2}), we obtain (see Appendix)
\begin{eqnarray}\label{mn-nm}
&&[z_m, z_n]=(a_m a_n^*-a_n a_m^*)\,a^2\,U(H,J^2)~,~~~~\mbox{where}\\
\nonumber &&U=X_N(H-2,J^2)\,\tilde Y_N(H,J^2)-Y_N(H-2,J^2)\,\tilde
X_N(H,J^2)~,
\end{eqnarray}
and due to (\ref{X,Y=1}), the commutator $[z_m, z_n]$ vanishes.

Calculating $[z_m^*, z_n]$ in a similar way we get the following
structure (see Appendix)
\begin{eqnarray}\label{m*n-nm*}
[z_m^*, z_n]=\delta_{mn}\,U_0+a_m^* a_n\, U_1+ a_n^* a_m\,U_2
+a_m^* a_n^*\,a^2\,U_3 +a_m a_n\,{a^*}\,^2\,U_4~,
\end{eqnarray}
where $U_0, ...,U_4$ are functions of the scalar variables $H$,
$J^2$ like $U$ in (\ref{mn-nm}). The exchange relations
(\ref{fa_n}) and (\ref{fa_n*}) provide $U_3=U_4=0,\,\,
U_2=-U_1=2,\,\, U_0=-2(H+\alpha)$ and this completes the proof of
(\ref{so1}).

The unitarity of our representations is guaranteed as long as
$z_n$ and $z_n^*$  expressed by (\ref{z_n^}) in terms of $a_n$ and $a_n^*$
are adjoint to each other. This in turn implies selfadjoint $X_N$ and $Y_N$,
hence positivity of the operator expression under the square root 
in  (\ref{Y_N=}). 
With the help of (\ref{M^2^}) this condition can be written as 
$a^*\,^2a^2+2(H+\alpha)(2\alpha-N+2)\geq 0$ and it reproduces the
unitarity bound (\ref{UB}). 

Irreducibility of the representations is certainly given as long as the
creation-annihilation operators $a_n^*$ and $a_n$ can be expressed
in terms of the symmetry generators. This requires an inversion of
(\ref{z_n^}).
As a first step for this inversion
we need the quantum analog of  (\ref{z^2=a^2}), which is obtained from
(\ref{z_n^}) and (\ref{z_n=2}) 
\begin{equation}\label{z^2=a^2^}
z^2=F_N(H, J^2)\,{a}^2~. 
\end{equation}
Based on this relation we get finally
\begin{equation}\label{a=z^}
a_n=\frac{1}{2(E+1)}\left(X_N\,z_n-\frac{Y_N}{F_N}\,{z}^2\,z_n^*\right)~,
~~~~~~~~~~
a_n^*=\frac{1}{2E}\left(z_n^*\,X_N-z_n\,{z^*}^2\,\,\frac{Y_N}{F_N}\right)~.
\end{equation}
This inversion formula is well-defined as long as $\alpha $ is above the
unitarity bound. But note also that for $\alpha $ just on the unitarity bound
$ F_N=\sqrt{{a^*}\,^2a^2}\,$,
which cannot be inverted within the full oscillator space. Therefore
irreducibility is lost at the unitarity bound. 

\section{Conclusions}

Our main result is the realization of spin zero $so(2,N)$ representations
in terms of just one set of $N$-dimensional oscillator operators. The
formulas defining this representation are (\ref{M_mn^}), (\ref{E^}),
(\ref{z_n^}) and (\ref{Y_N=})  (or equivalently (\ref{z_n=1}) and
(\ref{X,Y=})) and they reproduces the well-known unitarity bound (\ref{UB}).

The whole construction was based on an one to one map of the space of 
dynamical integrals of a scalar massive particle in $AdS_{N+1}$ 
to the phase space of a $N$-dimensional oscillator. 
This distinguishes our approach
from the oscillator like representations of non-compact groups,
developed in \cite{G}, where, using more oscillators, all symmetry generators
are represented bilinearly in the oscillators operators and the representation
space is selected out from the oscillator Fock space by some conditions, which
can be interpreted as coherent state constraints.

Our $so(2,N)$ representations arose as a by-product of scalar particle
dynamics. It would be interesting to extend these considerations to 
particles with spin and to construct the corresponding $so(2,N)$ 
representations with non-zero $so(N)$ weights.

In a forthcoming paper \cite{DJ3} we will analyze in detail the quantization of massless
particles. This will correspond to the special case $\alpha =(N\pm 1)/2$
in which the symmetry group is enlarged to the conformal group of
$AdS_{N+1}$. 
There we further comment on the reducibility of our representation of 
$so(2,N)$ at the unitarity bound and discuss its relation 
to the singleton representations \cite{F} and to massless
particle dynamics in one dimension less, i.e. in $AdS_{N}$.

\vspace{3mm}

\noindent
{\bf Acknowledgments.~}
\noindent
We thank H. Nicolai for discussions.
G.J. is grateful to Humboldt University and AEI Potsdam for
hospitality. His research was supported by grants from the DFG, GRDF
and GAS. H.D. was supported in part by DFG with the grant DO 447-3/3.

\vspace{0.3cm}

\setcounter{equation}{0}

\def\theequation{A.\arabic{equation}}

{\bf \large{Appendix}}

\vspace{0.1cm}

\noindent Let us consider the operators $H_N$ and $ J_N^2$ given
by (\ref{X,Y=}), (\ref{M^2^}) and calculate their exchange
relations with the operators $a_n$ and $ b_n= a^2 a_n^*$. The
canonical commutators (\ref{CCR}) provide
\begin{eqnarray}\label{H,a_n}
&&H_N \,a_n=a_n\,(H_N-1)~,~~~~J_N^2\,a_n =a_n(J_N^2-2H_N+1)+2b_n~,\\
\label{J,a_n} &&H_N \,b_n=b_n\,(H_N-1)~,~~~~J_N^2\,b_n
=b_n(J_N^2+2H_N+1)-2a_n(H_N^2-J_N^2)~.
\end{eqnarray}
These relations are diagonalized by the operators
\begin{equation}\label{c_n}
 c_n= b_n-a_n(H_N+J_N),~~~~~~~d_n= b_n-a_n(H_N-J_N),
 \end{equation}
in the following form
\begin{eqnarray}\label{H,c_n}
&&H_N \,c_n=c_n\,(H_N-1)~,~~~~~J_N^2\,c_n =c_n(J_N-1)^2~,\\
\label{J,c_n} &&H_N \,d_n=d_n\,(H_N-1)~,~~~~J_N^2\,d_n
=d_n(J_N+1)^2~.
\end{eqnarray}
Here $J_N=\sqrt{J_N^2}\,$ and the square root from the positive
operator is defined in a standard way.
Introducing new scalar variables
\begin{equation}\label{u,v}
u=H_N+J_N~~~~~~~\mbox{and}~~~~~~v=H_N-J_N~,
\end{equation}
the functions  (\ref{Y_N=}) can be written as
\begin{equation}\label{X(u,v)}
X_N=\frac{u\sqrt{u+\beta}-v\sqrt{v+\beta}}{u-v}~,~~~~
Y_N=-\frac{\sqrt{u+\beta}-\sqrt{v+\beta}}{u-v}~,~~~
\mbox{with}~~\beta=2\alpha-N~.
\end{equation}
Due to (\ref{H,c_n})-(\ref{J,c_n}), a function $f(u,v)$ 
satisfies the exchange relations
\begin{eqnarray}\label{f,c_n1}
f(u, v)\,c_n =\,c_n f(u-2,v)~,~~~~~ f(u, v)\,d_n =\,d_n f(u,v-2)~.
\end{eqnarray}
Inverting (\ref{c_n})
\begin{equation}\label{a_n=c_n}
 a_n= d_n\frac{1}{u-v}-c_n\,\frac{1}{u-v}~,~~~~~
 b_n=d_n\frac{u}{u-v}-c_n\frac{v}{u-v}~,
\end{equation}
and using (\ref{f,c_n1}), we obtain
\begin{eqnarray}\label{fa_n}
 &&f(u,v)\,a_n=a_n\,\frac{u f(u-2,v)-v f(u,v-2)}{u-v}
-b_n\,\frac{f(u-2,v)-f(u,v-2)}{u-v}~,\\
\nonumber &&f(u,v)\,b_n =a_n\,\frac{uv f(u-2,v)-uv f(u,v-2)}{u-v}
-b_n\,\frac{v f(u-2,v)-u f(u,v-2)}{u-v}~.
\end{eqnarray}
Applying these exchange relations to the functions (\ref{X(u,v)})
we get
\begin{eqnarray}\label{X,a_n}
 &&X_N\,a_n=a_n\,X_{(1)}+b_nY_{(1)}~,~~~
 Y_N\,b_n=a_n\,X_{(2)}+b_nY_{(2)}~,\\\label{X,a_n1}
&& X_N\,b_n=a_n\,X_{(3)}+b_nY_{(3)}~,~~~
 Y_N\,a_n=a_n\,X_{(4)}+b_nY_{(4)}~,~~~
\end{eqnarray}
where
\begin{eqnarray}\nonumber
&&X_{(1)}=
\frac{u(u-2)\sqrt{u-2+\beta}-uv\sqrt{v+\beta}}{(u-v)(u-2-v)}-
\frac{uv\sqrt{u+\beta}-v(v-2)\sqrt{v-2+\beta}}{(u-v)(u-v+2)}\,,
\\\label{Y_1}
&&Y_{(1)}=-\frac{(u-2)\sqrt{u-2+\beta}-v\sqrt{v+\beta}}{(u-v)(u-2-v)}+
\frac{u\sqrt{u+\beta}-(v-2)\sqrt{v-2+\beta}}{(u-v)(u-v+2)}~;
\end{eqnarray}
\begin{eqnarray}
\nonumber
&&X_{(2)}=-\frac{uv\sqrt{u-2+\beta}-uv\sqrt{v+\beta}}{(u-v)(u-2-v)}+
\frac{uv\sqrt{u+\beta}-uv\sqrt{v-2+\beta}}{(u-v)(u-v+2)}~,
~~~~~~~~~~~~\\
\label{Y_2}
&&Y_{(2)}=\frac{v\sqrt{u-2+\beta}-v\sqrt{v+\beta}}{(u-v)(u-2-v)}-
\frac{u\sqrt{u+\beta}-u\sqrt{v-2+\beta}}{(u-v)(u-v+2)}~.
\end{eqnarray}
and the functions $X_{(3)}$, $Y_{(3)}$, $X_{(4)}$, $Y_{(4)}$ are
expressed in a similar way. Since $b_n=a_n^*a^2+2a_n$, the
operator $z_n=X_Na_n+Y_Nb_n$ can be rewritten  in the form
(\ref{z_n=2}) with
\begin{eqnarray}\label{tildeX}
\tilde X_N=X_{(1)}+X_{(2)}+2(Y_{(1)}+Y_{(2)})~,~~~\tilde
Y_N=Y_{(1)}+Y_{(2)}~.
\end{eqnarray}
Then, Eqs. (\ref{Y_1})-(\ref{Y_2}) yield
\begin{eqnarray}\label{tildeX1}
&&\tilde X_N=\frac{(u-2)\sqrt{u-2+\beta}-(v-2)\sqrt{v-2+\beta}}{u-v}~,\\
\label{tildeY} &&\tilde
Y_N=-\frac{\sqrt{u-2+\beta}-\sqrt{v-2+\beta}}{u-v}~.
\end{eqnarray}
and passing back from $\,(u, v)\,$ to  $\,(H,J)\,$, we arrive at
(\ref{X',Y'}).

The commutator $[z_m,z_n]$ is obtained by anti-symmetrization of
\begin{eqnarray}\label{mn}
z_m\,z_n=X_Na_ma_n\tilde X_N+X_Na_ma_n^*a^2\tilde
Y_N+Y_Na^2a_m^*a_n\tilde X_N+Y_Na^2a_m^*a_n^*a^2\tilde Y_N~,
\end{eqnarray}
where the representations (\ref{z_n^}) and (\ref{z_n=2}) are used
for $z_m$ and $z_n$ respectively. Taking into account that the
operator $J_{mn}=i(a_n^*a_m -a_m^*a_n)$ commutes with scalar
operators and that $X_N(H,J^2)a^2=a^2X_N(H-2,J^2)$, the commutator
$[z_m,z_n]$ reduces to (\ref{mn-nm}).

To represent the commutator $[z_m^*,z_n]$ in the form
(\ref{m*n-nm*}) we also use the exchange relations of $a_n^*$ and
$b_n^*$ with scalar operators
\begin{eqnarray}\label{fa_n*}
&&f(u,v)\,a_n^*=a_n^*\,\frac{u f(u+2,v)-v f(u,v+2)}{u-v}
-b_n^*\,\frac{f(u+2,v)-f(u,v+2)}{u-v}~,\\
\nonumber &&f(u,v)\,b_n^* =a_n^*\,\frac{uv f(u+2,v)-uv
f(u,v+2)}{u-v} -b_n^*\,\frac{v f(u+2,v)-u f(u,v+2)}{u-v}~,
\end{eqnarray}
which can be derived similarly to (\ref{fa_n}).
Writing the terms $z_m^*z_n$ and $z_n z_m^*$ as
\begin{eqnarray}\label{m*n}
&&z_m^*z_n=(a_m^*X_N+b_m^*Y_N)(a_n(\tilde X_N-2\tilde Y_N)
+b_n\tilde Y_N)~,~~~~\mbox{and}\\ \nonumber &&z_n z_m^*=(a_n\tilde
X_N +a_n^*a^2\tilde Y_N)(a_m^*X_N+b_m^*Y_N)
\end{eqnarray}
respectively, and applying (\ref{fa_n}), (\ref{fa_n*}) and then
(\ref{CCR}), we obtain the commutator $[z_m^*,z_n]$ in the form
(\ref{m*n-nm*}).

\end{document}